\documentclass[conference]{IEEEtran}

\usepackage[utf8]{inputenc}
\usepackage[T1]{fontenc}
\usepackage{tabularx}
\usepackage{placeins}
\usepackage{algorithm2e}
\usepackage{amsmath}
\usepackage{graphicx}
\usepackage{comment}
\usepackage{listings}

\usepackage{float} 

\includecomment{comment} 
\DeclareUnicodeCharacter{2264}{<=}
\begin{document}

\title{Deriving ChaCha20 Key Streams From Targeted Memory Analysis}

\author{\IEEEauthorblockN{Peter McLaren, William J Buchanan, Gordon Russell, Zhiyuan Tan}
\IEEEauthorblockA{School of Computing,\\
Edinburgh Napier University,\\
Edinburgh, UK.}
}


\maketitle

\begin{abstract}
There can be performance and vulnerability concerns with block ciphers, thus stream ciphers can used as an alternative. Although many symmetric key stream ciphers are fairly resistant to side-channel attacks, cryptographic artefacts may exist in memory. This paper identifies a significant vulnerability within OpenSSH and OpenSSL and which involves the discovery of cryptographic artefacts used within the ChaCha20 cipher. This can allow for the cracking of tunneled data using a single targeted memory extraction. With this, law enforcement agencies and/or malicious agents could use the vulnerability to take copies of the encryption keys used for each tunnelled connection. The user of a virtual machine would not be alerted to the capturing of the encryption key, as the method runs from an extraction of the running memory. Methods of mitigation include making cryptographic artefacts difficult to discover and limiting memory access.
\end{abstract}


\section{Introduction}

There is an increasing challenge between the rights of citizens to privacy and the rights of society to protect itself from adversaries \cite{bauman2014after}\cite{Iphofen}. The breaking of encryption tunnels is thus one of the major debating points, and where law enforcement agencies often aim to gather tools and methods which break these tunnels, or where we fix vulnerabilities in tools in order to avoid these tunnels from being broken. In most cases we now perform a key negotiation phase – typically with ECDH (Elliptic Curve Diffie-Hellman) – and then use a symmetric key method to encrypt the traffic within the tunnel. 

The cracking of the key exchange process and of the symmetric key used in the tunnel are, in most cases, too costly to crack. Unfortunately, the key exchange process can leave behind trails of evidence in memory which can provide significant clues to the symmetric key being used. While this has been demonstrated for block ciphers, such as for the Advanced Encryption Standard (AES) \cite{halderman2009lest}\cite{maartmann2009persistence}, this paper outlines how well-used applications such as OpenSSH and OpenSSL allow for every generated key in the ChaCha20 stream cipher to be revealed within a fairly fast discovery time. As virtualized environments enable access to virtual machine resources from more privileged levels such as hypervisors or hypervisor consoles, applications operating at that level can extract live virtual machine memory. Extraction is most effective when a virtual machine is paused but it is not necessary. So, virtualized environments present an opportunity to find keys without impacting the target and for target applications to be unaware of extraction.

The rest of the paper is structured as follows. Section II discusses related research including side-channel studies and background on stream ciphers and ChaCha20 cipher implementations is presented in Section III. Section IV provides relevant details of the framework and its implementation is given in Section V. The results are presented and discussed in Section VI and conclusions drawn in Section VII.

\section{Related Work}
This paper focuses on the decrypting network traffic encrypted with ChaCha20-Poly1305 cipher. Prior studies have investigated potential vulnerabilities in cipher design and in cipher implementation. Researchers have found no vulnerabilities in ChaCha20 design. For example, differential attacks using techniques such as identifying significant key bits only succeeded with reduced cipher rounds and significant volumes of plaintext-ciphertext pairs \cite{aumasson2008new}\cite{maitra2016chosen}. Combined linear and differential analysis improves performance, but is similarly restricted \cite{choudhuri2016differential}. 

ChaCha20 implementations may be vulnerable to side-channel attacks. While the cipher design may prevent timing attacks \cite{bernstein2008salsa20}, correlating power electromagnetic radiation when specific cryptographic activities are performed may leak key stream information \cite{jungk2017don}\cite{adomnicai2017bricklayer}. Engendering instruction skips, for example by using a laser or electromagnetic pulse, could potentially produce the key stream but timing the activity would be challenging \cite{kddi}. Furthermore, these approaches may be impractical in real-world scenarios. 

Cryptographic artefacts have been found in device memory. For instance, RSA keys may be discovered in virtual machine images \cite{Rocha2011} \cite{Saxon2015}. Studies have also discovered DES and AES cipher keys in cold-boot attacks \cite{halderman2009lest}, Skipjack and Twofish key blocks in virtual memory \cite{maartmann2009persistence}, and AES session keys in virtual memory \cite{taubmann2016tlskex}. Although these approaches use entropy measures to determine possible keys, they do not decrypt ciphertext encrypted with ciphers such as AES in Counter mode and ChaCha20 which require nonces/initialization vectors. This study builds on the TLSkex  \cite{taubmann2016tlskex} and MemDecrypt studies \cite{mclarenssh2019} which used privileged monitors to extract identified virtual machine process memory to identify TLS 1.2 AES keys, and SSH AES keys and initialization vectors, respectively. Instead, this study uses a different algorithm to find ChaCha20 cipher keys and nonces in device memory enabling SSH and TLS sessions to be decrypted in a non-invasive manner. The approach may enable decryption of Adiantum encrypts \cite{crowley2018adiantum}, the Google disk-encryption algorithm based on XChaCha20, an extension to ChaCha20 and Salsa20 with a longer nonce \cite{bernstein2011extending}.

\section{Stream ciphers}
Secure protocols use encryption to provide confidentiality for secure communications between parties. While asymmetric encryption ciphers are used in secure protocol set-up stages, for performance symmetric ciphers encrypt the confidential information. Symmetric ciphers are commonly classified as being stream ciphers, where plaintext is encrypted bit-by-bit or byte-by-byte or block ciphers, where blocks of a specific size are encrypted. This paper focuses on stream ciphers.

Stream ciphers generate a random key stream from an evolving state \cite{biryukov2004block}. The key stream is then typically XOR-ed with the plaintext to generate ciphertext \cite{biryukov2004block}. Software implementations for stream cipher have been found to be faster than block ciphers although possibly more difficult to implement \cite{klein2013stream}. Stream ciphers have typically been used in embedded technologies such as Internet of Things (IoT) devices and smartphones \cite{manifavas2016survey}. Stream ciphers, as well as block ciphers, have been supported by secure protocols. However, with vulnerabilities leading to the planned deprecation of the RC4 stream cipher for protocols such as Secure Shell (SSH) \cite{Camara} and Transport Layer Security (TLS) \cite{popova}, alternative stream ciphers have been under consideration. In particular, the ChaCha20 stream cipher with the Poly1305 authenticator \cite{nir} has been adopted in secure protocol implementations such as OpenSSH \cite{OpenSSH}, and OpenSSL \cite{OpenSSL}, as well for Google Chrome on smartphones \cite{Ianix}. The discovery of the key stream or inputs to key stream generation may be an unacceptable vulnerability for stream ciphers including ChaCha20-Poly1305. 

Stream cipher cryptographic artefacts are memory-resident at points in time. Furthermore, these artefacts, as well as the key stream are on a program stack, in the heap or in shared memory. If timely acquisition of target device memory is obtained, the stream or artefacts may be discovered. With the growth of forensic technologies that enable memory access to targets such as desktops, servers, smartphones, and IoT devices, an opportunity exists to discover these artefacts. Although researchers have investigated ChaCha20 side-channel attacks with intermediate state leakage \cite{jungk2017don} and without \cite{adomnicai2017bricklayer}, no studies analyze memory so this paper presents a new approach to decrypting secure communications encrypted with the ChaCha20 cipher. A single memory extract suffices to find the cryptographic artefacts for decryption. Furthermore, the approach is faster and less invasive than methods that use side-channel techniques.

OpenSSH and PuTTY implement chacha20-poly1305@openssh.com while OpenSSL \cite{Ianix} versions above 1.1 implement variations where the asymmetric cipher used in key exchange vary, e.g. ECDHE-RSA-CHACHA20-POLY1305 \cite{langley2016chacha20}. Both implementations adhere to the RFC \emph{ChaCha20 and Poly1305 for IETF Protocols} \cite{nir} which is based on Bernstein’s ChaCha20 cipher proposal \cite{bernstein2008chacha}, a variation of the earlier Salsa20 cipher \cite{bernstein2008salsa20}. 

\subsection{ChaCha20 ciphering process}
The ChaCha20 cipher generates key streams of 64-bytes. Inputs to key stream generation are independent of plaintext or ciphertext similar to other eSTREAM proposals \cite{Hutchison} but unlike stream ciphers such as Helix \cite{bernstein2008salsa20}. This enables parallel ciphertext generation with consequent performance improvement. 20 rounds of mathematical calculations using XOR, addition and rotation using as inputs four 4-byte constants, a random 32-byte key, a 4-byte counter, and a 12-byte nonce (Bernstein originally specified the nonce and counter lengths to be eight). However, this is not a material difference in the investigations). The 4-byte constants are 0x61707865, 0x3320646e, 0x79622d32, and 0x6b206574 or in ASCII ‘apxe’, ‘3 dn’, ‘yb-2’, and ‘k et’. In ChaCha20 these strings are concatenated. The counter, which typically starts at 0 or 1 increments for each 64-byte plaintext block \cite{nir}. 

In the chacha20-poly1305@openssh.com implementation, memory is allocated to hold a structure comprising key stream input fields, the 64-byte key stream itself, and an index pointer. The packet lengths are encrypted separately from the remainder of the payload so four structures are required: two for encrypting outgoing lengths and payloads, and the same again for incoming encrypted data. The memory contents of the concatenated constant string are \emph{expand 32-byte k}. The nonce is a sequence number and counters for the encrypted packet lengths and payloads are zero and unity, respectively.

The OpenSSL implementation of ChaCha20-Poly1305 differs in a number of respects as required by IETF RFC ChaCha20-Poly1305 Cipher Suites for Transport Layer Security (TLS) \cite{langley2016chacha20}. As encrypting packet length is not required in TLS two memory structures are used for encrypting outgoing messages and decrypting incoming messages. Memory used for ChaCha20 inputs is temporary as the encryption structure is assembled from other sources. Also, the nonce is an XOR of the sequence number and the vector generated during the initial handshake when the cipher keys are obtained. 
 \begin{figure*}[!htb]
        \center{\includegraphics[width=0.8\textwidth]
        {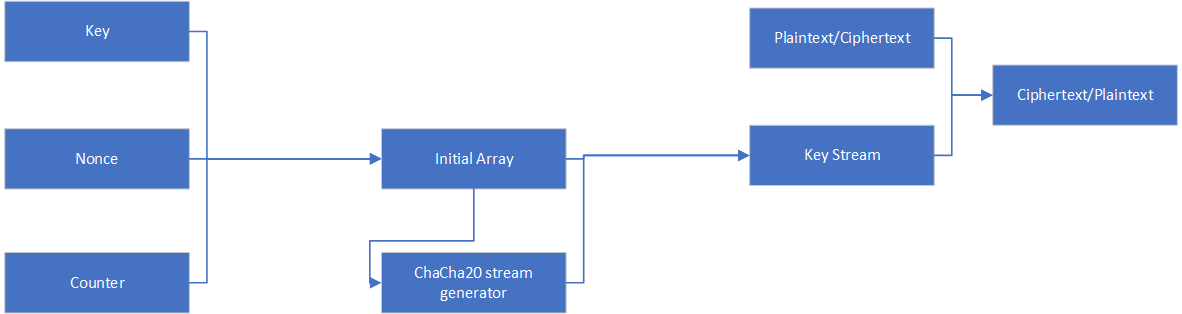}}
        \caption{\label{fig:ChaCha}ChaCha Encryption/Decryption Flow}
      \end{figure*}    
      
ChaCha20 takes a 256-bit key and a 32-bit nonce (and which includes a counter). This creates a key stream which is then XOR-ed with the plaintext stream as illustrated in Figure \ref{fig:ChaCha}. In software-only implementations, it is often more than three times faster than AES \cite{langley2016chacha20}, and is well suited to lower-powered devices and in real-time communications. ChaCha20 operates on 32-bit words at a time with a key of 256 bits (K=($k_0$, $k_1$, $k_2$, $k_3$, $k_4$, $k_5$, $k_6$, $k_7$). This outputs blocks of 512 bits for the key stream (Z), and which is XOR-ed with the plaintext stream. The state of the encryption is stored within 16x32-bit word values and arranged as a 4x4 matrix:
\begin{equation}
\begin{bmatrix}
x_0 & x_1 & x_2 & x_3\\
x_4 & x_5 & x_6 & x_7\\
x_8 & x_9 & x_{10} & x_{11}\\
x_{12} & x_{13} & x_{14} & x_{15}\\
\end{bmatrix}
\end{equation}

The initial state contains 16x32-bit values with constant values (0x61707865, 0x3320646e, 0x79622d32, 0x6b206574) the key ($k_0$, $k_1$, $k_2$, $k_3$, $k_4$, $k_5$, $k_6$, $k_7$), the counter ($c_0$) and the nonce ($n_0$,$n_1$,$n_2$,$n_3$):
 \begin{equation}
\begin{bmatrix}
\scriptscriptstyle 0x61707865 & \scriptscriptstyle 0x3320646e & \scriptscriptstyle 0x79622d32b & \scriptscriptstyle 0x6b206574\\
k_0 & k_1 & k_2 & k_3\\
k_4 & k_5 & k_{6} & k_{7}\\
c_{0} & n_{0} & n_{1} & n_{2}\\
\end{bmatrix}
\end{equation}

The counter thus has 32-bits (1 x 32 bits), and the nonce has 96-bits (3 x 32 bits). ChaCha20 then defines a quarter round as shown in Algorithm \ref{alg:ChaChaQuarterRound}.

\begin{algorithm}
\SetAlgoLined
\KwResult{QR(a,b,c,d)}
$a = a + b$\;
$d = d \oplus a$\;
$d = (d)<<16$\;
$c = c + d$\;
$b = b \oplus c$\;
$b = (b)<<12$\;
$a = a + b$\;		
$d = d \oplus a$\;
$d = (d)<<8$\;
$c = c + d$\;
$b = b \oplus c$\;
$b = (b)<<7$\;
\caption{Quarter round}
\label{alg:ChaChaQuarterRound}
\end{algorithm}

There are then 20 rounds (10 for column rounds and 10 for diagonal rounds) as shown by Algorithm \ref{alg:ChaChaKeyStream}.

\begin{algorithm}
\SetAlgoLined
\KwData{X is created with K, c and n}
\KwResult{Z is the resultant key stream.}
$y \longleftarrow  X$\;
\For {i = 0 to 9} {
 $(x0, x4, x8, x12) \longleftarrow  QR(x0, x4, x8, x12)$\;
 $(x5, x9, x13, x1) \longleftarrow  QR(x5, x9, x13, x1)$\;
 $(x10, x14, x2, x6) \longleftarrow  QR(x10, x14, x2, x6)$\;
 $(x15, x3, x7, x11) \longleftarrow  QR(x15, x3, x7, x11)$\;
 $(x0, x5, x10, x15) \longleftarrow  QR(x0, x5, x10, x15)$\;
 $(x1, x6, x11, x12) \longleftarrow  QR(x1, x6, x11, x12)$\;
 $(x2, x7, x8, x13) \longleftarrow  QR(x2, x7, x8, x13)$\;
 $(x3, x4, x9, x14) \longleftarrow  QR(x3, x4, x9, x14)$\;
}
$ Z \longleftarrow  X + y\; $
\caption{Keystream}
\label{alg:ChaChaKeyStream}
\end{algorithm}

\section{Decryption Framework}
The decryption framework is comprised of data capture, analysis, and decrypt components. Each component is modifiable or replaceable so that different devices, target operating systems, ciphers, and protocols can be addressed. Details of the component design for ChaCha20 are presented in the following paragraphs. 

\textbf{Data Collection}. Target device network packets and volatile memory are extracted. Complete SSH and TLS sessions originating from the target device are captured for later examination. The indicator for extraction of target volatile is the client transmission of a protocol message indicating that the initialization phase is complete. These messages are \emph{New Keys} and \emph{Client Finished} for SSH and TLS Version 1.2 respectively. Cryptographic artefacts have then been generated and are likely to be memory-resident and memory can be extracted for any outgoing message in the network session.

\textbf{Memory Analysis}. Candidate cryptographic artefacts are discovered in memory extracts. Initially, the component searches for the constant string \emph{expand 32-byte k} to discover candidate ChaCha20 data structures. Although unlikely for the string to be present in non-base structures, a second step assesses whether the 32-byte block after the constant string in a candidate base structure’s might be a key. Encryption keys must be unpredictable, i.e. random. Key randomness can be evaluated using the Shannon entropy definition \cite{shannon1948mathematical}:\\
\begin{equation}
H=-\sum_{i=1}^{n} p(i) \log_2 p(i)
\end{equation}

where $p_i$ is the normalized frequency of the $i$th byte in the message i.e. $p(i)=f(i)/n$. So, where the entropy of the 32-byte block exceeds a threshold, a candidate base structure is identified and the key, nonce and counter are retained.

\textbf{Decrypt Analysis}. Cryptographic artefacts output by the memory analysis component are input parameters in decrypt analysis. In each instance, the candidate key and nonce group is decrypted and verified according to the protocol used. For SSH, the groups are used to decrypt incoming and outgoing packet lengths and payloads. The packet length groups are identified if the decrypted packet length meets Equation (2) for short packets, typically in the authentication and channel set-up phases. For larger packets, the decrypted packet length supports SSH packet reassembly. Decrypts with the remaining cryptographic groups are analyzed to establish compliance with SSH protocol specifications. For TLS, decrypts are analyzed to establish compliance with TLS. protocol specifications.

\begin{equation}
\begin{split}
packet\ data\ length\ = \\ 
     & decrypted\ packet\ length\ + \\
     & size(packet\ length field)\ + \\
     & size(MAC field)
\end{split}
\end{equation}

\section{Implementation}
\begin{figure*}[!htb]
        \center{\includegraphics[width=0.6\textwidth]
        {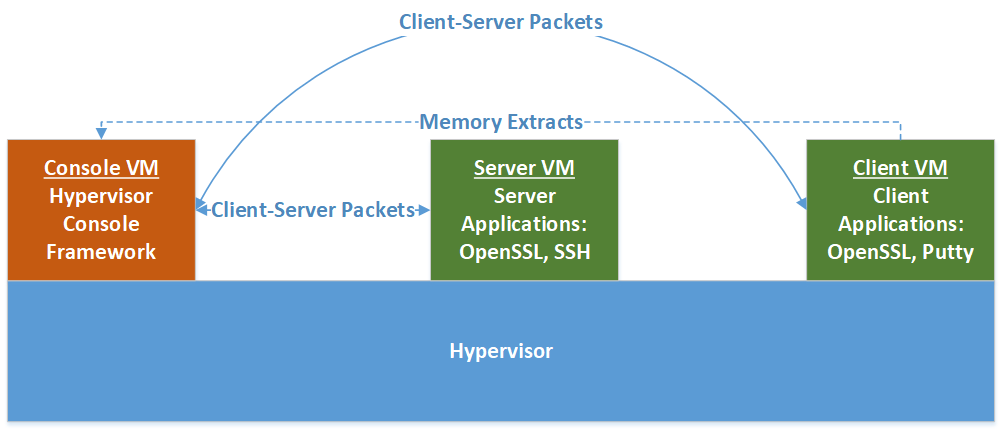}}
        \caption{\label{fig:Framework} Virtualized environment framework}
      \end{figure*}

The framework is implemented in a virtualized environment as illustrated in Figure \ref{fig:Framework} . Implementations on other technologies which facilitate packet capture and target memory access should be possible. The Xen hypervisor \cite{Xen} offers benefits over alternatives including the presence of LibVMI \cite{LibVMI} and PyVMI \cite{Payne} libraries providing access to the volatile memory of live virtual machines. Because of its small trusted computing base, Xen is managed by a privileged virtual machine, which runs or initiates the framework components. The privileged virtual machine also provides the unprivileged guests with network and disk access. 

The data collection component inspects virtual machine network traffic.  Each packet is redirected to a local queue using an iptables rule and NetFilterQueue 0.8.1 \cite{Kerkhoff}, and protocol fields to determine whether SSH or TLS sessions have been initiated. Each SSH and TLS packet is written to file. Memory extraction uses PyVMI, LibVMI, and Volatility libraries \cite{volatilityfoundation}.

Memory analysis obtains candidate ChaCha20 cryptographic artefacts. Each memory extract file is searched for the constant string. Although false positives are unlikely, additionally the entropy of the following 32-byte block is compared with a threshold, experimentally found to be 4.5. If the threshold is exceeded, a base structure has been identified and the key, nonce, and counter fields are retained for decryption as groups. 

\begin{algorithm}
\SetAlgoLined
\KwData{Extracts folder, entropy threshold}
\KwResult{Z = candidate artefacts}
\For {extract in folder} {
    i = 0\;
	\While {not extract EOF} {
		i :=  locate ‘expand 32-byte k’ in extract\;
		\If {i > 0} {
		    \eIf {entropy extract[i+16:i+48] > threshold} {
			Z += key, nonce, and counter\;
			i += 64\;
			}{i += 16\;}
		}
	}
}
\caption{Memory Analysis}
\end{algorithm}

The decrypt analysis component iterates through the key and nonce groupings. For SSH, the first four encrypted bytes in packets are decrypted using the Chacha20poly1305 package with counter value 0 \cite{Chacha20poly1305}. For a valid group the first four bytes represent the packet length and Equation (2) holds. For ChaCha20, the decrypts with other groups are validated. If the decrypted padding length obeys Equation (3) as specified in SSH Transport Layer Protocol \cite{ylonen2005secure} the following blocks are decrypted to evaluate compliance with the SSH authentication and channel set-up specifications \cite{ylonen2005auth}\cite{ylonen2005connectoin}. For TLS, the entire data block is decrypted with cryptographic artefact groups. Although TLS supports other higher-level protocols such as SMTP, HTTP-over-TLS is perhaps the most commonly use and so compliance of the decrypt with the HTTP 1.1 specification \cite{fielding1999hypertext} is assessed. Valid decrypts are retained for user inspection.\\
\begin{equation}
4 \leq padding\ length \leq 255
\end{equation}

The physical environment for experimentation is a Core 2 Duo Dell personal computer with 40\,GB of disk storage and 3\,GB of RAM. This hosts the hypervisor, the privileged virtual machine, an unprivileged Windows virtual machine, and an unprivileged Ubuntu virtual machine. The hypervisor is Xen Project 4.4.1 and the hypervisor console is Debian release 3.16.0-4-amd64 Version 1. Tests run on untrusted Windows client and Linux server virtual machines. The Windows clients runs the Windows 10 (10.0.16299) operating system with 2\,GB of memory and 40 GB of disk. A Linux virtual machine runs an Ubuntu 14.04 build (“Trusty”) with 512\,MB of allocated memory and 4\,GB of disk storage. The PuTTY suite \cite{PuTTY} is used for SSH client application testing with SSH server functionality provided by openssh-server. OpenSSL 1.1.0h provides TLS client and server functionality.

\section{Evaluation}
For SSH evaluation the PuTTY ‘pscp’ program is executed from the Windows command line using requests of the form:\newline

\textit{pscp -P nnnn 'filename' @ipaddress:/home/name}\newline

where \textit{nnnn} is the target port, \textit{‘filename’} is the file being transmitted, \textit{name} is a user account on the target Ubuntu server, \textit{ipaddress} is the target server IP address and \textit{/home/name} is the Ubuntu server target folder for the transmitted file. An Ubuntu service is started from the bash command line to listen to client SSH messages with requests of the form: \newline

\textit{/usr/sbin/sshd -f /root/sshd\_config -d -p nnnn }\newline

where \textit{nnnn} is the port number the service listens on and \textit{sshd\_config} contains configuration details. Here, \textit{sshd\_config} contains the string ‘ciphers chacha20-poly1305@openssh.com’. Client and server tests execute OpenSSL from the command line. The OpenSSL server emulates a web server with:\newline

\textit{openssl s\_server -accept 443 -cert crt.pem -key key.pem -WWW }\newline

where \textit{crt.pem} and \textit{key.pem} are certificate and key files. The client connects to the OpenSSL server with: \newline

\textit{openssl s\_client [-cipher CIPHER] -connect a.b.c.d:443}\newline

where \textit{CIPHER} identifies the encryption cipher, key exchange and authentication algorithms (here, ECDHE-RSA-CHACHA20-POLY1305) and \textit{a.b.c.d} the OpenSSL server IP address. Client input simulates browser requests, e.g. ‘GET / HTTP/1.1’, ‘Host: a.b.c.d.’, Accept-Encoding: gzip, deflate’, Accept: */*.

\begin{figure*}[!htb]
        \center{\includegraphics[width=0.6\textwidth]
        {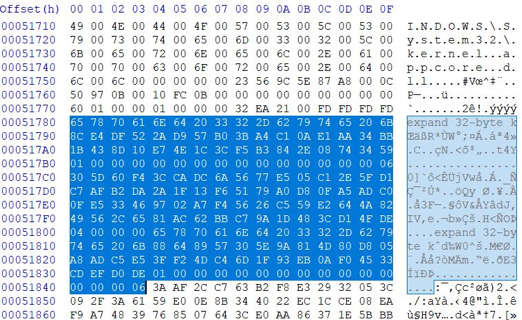}}
        \caption{\label{fig:SSHbase}SSH Base Structures in Memory}
      \end{figure*}

\begin{figure*}[!htb]
        \center{\includegraphics[width=0.6\textwidth]
        {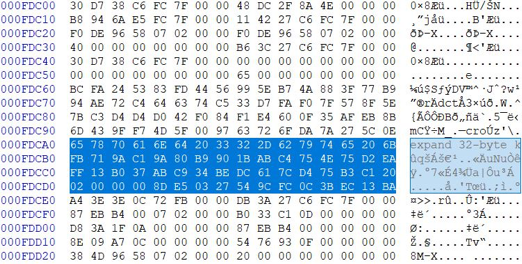}}
        \caption{\label{fig:TLSbase}TLS Base Structure in Memory}
      \end{figure*}

\begin{figure*}[!htb]
        \center{\includegraphics[width=0.5\textwidth]
        {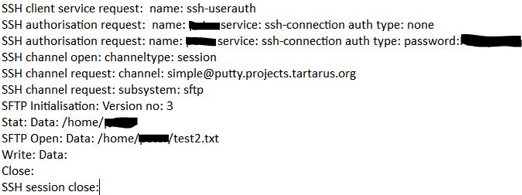}}
        \caption{\label{fig:SSHdecrypt}SSH Client Decrypt Example}
      \end{figure*}

\begin{figure*}[!htb]
        \center{\includegraphics[width=0.5\textwidth]
        {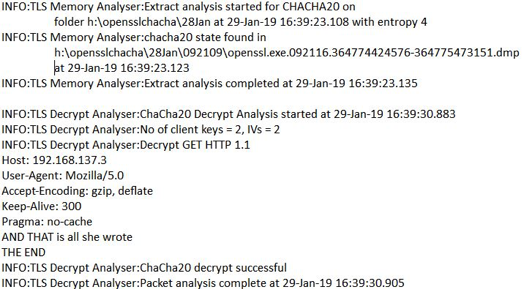}}
        \caption{\label{fig:MemDecrypt}TLS memory and Decrypt Analysis Output}
      \end{figure*}

ChaCha20 base structures were found in 100\% of instances. For the SSH chacha20-poly1305@openssh.com protocol four base stuctures were discovered in Windows and Linux application memory. For OpenSSL, one base structure was discovered for ECDHE-RSA-CHACHA20-POLY1305 as shown in the highlighted section of Figures \ref{fig:SSHbase} and \ref{fig:TLSbase} respectively. These discoveries leads to key stream generation and the rapid decryption of complete SSH sessions including server user credentials, file names and uploaded file contents between 150 bytes and 1 MB as well as the decryption of outgoing client TLS traffic. Differences between SSH and TLS relates to their implementations. PuTTY/OpenSSH structures are heap-resident so memory extracts for successful decryption may not be linked to client SSH message transmission. By contrast, OpenSSL structures are stack-resident and may therefore be overwritten. This limitation prevents Linux server successful searches for the constant string although high-entropy regions identify candidate encryption keys. Consequently, full SSH sessions are decryptable whereas only outgoing TLS sessions are currently decryptable when applying the algorithms. A SSH decrypt is illustrated in Figure \ref{fig:SSHdecrypt} and example output from the analysis stages is presented in Figure \ref{fig:MemDecrypt}.
 
Memory and decrypt analysis components decrypt rapidly. The memory analysis durations are less than 0.5 (SSH) and 0.1 (TLS) seconds as shown in Table \ref{tab:my_label}. As single extract analysis requires a maximum 0.018 seconds, parallel processing memory extract analysis offers significant performance opportunities. Although memory analysis is independent of file size, decrypt analysis durations are proportionate to the volume of encrypted traffic.
\begin{table}[h]
    \centering
    \caption{Memory Analysis Durations}
    \begin{tabular}{c|c|c}
    \hline
    &SSH&TLS\\
    \hline
         Maximum&0.407&0.027\\
         Minimum&0.007&0.011\\
         Mean&0.144&0.021\\
         Standard Deviation&0.153 &0.006\\
         \hline
    \end{tabular}
    \label{tab:my_label}
\end{table}
\section{Countermeasures}
 Fortunately, countermeasures to discovering the ChaCha20 basic structures exist, and hiding the constant string makes discovery more challenging. Possible measures are copying the constant string segments to registers and assembling the structure in the encryption routine, encrypting the constant string, or randomly segmenting the constant string. Perhaps, the most effective approach is assembling the base structure on the stack, as for OpenSSL, and clearing stack contents immediately after the encryption process. The cryptographic artefacts can still be discovered by searching for high-entropy measures but the process is comparatively slower with durations exceeding 4 minutes.
 
\section{Conclusions}
Implementations of ChaCha20-Poly1305 encryption using commonly used applications and libraries for SSH and TLS communications are vulnerable to decrypt analysis on a single memory extract. As memory analysis identifies cryptographic artefacts with 100\% success, the artefacts could be retained with network sessions for later decryption. This may benefit entities, such as cloud vendors, to assist state agencies in decrypting criminals' communications, without conflicting with local privacy laws. To achieve this aim, future work should focus on performance improvements such as multi-threading and pipe-lining, as well as investigating other protocols, encryption ciphers and modes, and technologies that uses encrypted communications channels.

\bibliographystyle{IEEEtran}
\bibliography{main}

\end{document}